\documentclass{book}
\usepackage{maxent,xspace}

\title{A FULL BAYESIAN APPROACH FOR INVERSE PROBLEMS}
\author{
{\sc Ali Mohammad-Djafari} \\ 
{\em Laboratoire des Signaux et Syst\`emes (CNRS-ESE-UPS)} \\  
{\em \'Ecole Sup\'erieure d'\'Electricit\'e,} \\ 
{\em Plateau de Moulon, 91192 Gif-sur-Yvette, France.} \\ 
{\em E-mail: djafari@lss.supelec.fr}
}
\def\bm#1{\mbox{\boldmath $#1$}}
\def\xb{\bm{x}}
\def\yb{\bm{y}}

\def\nb{\bm{n}}
\def\gb{\bm{g}}

\def\Ab{\bm{A}}
\def\Db{\bm{D}}
\def\Mb{\bm{M}}
\def\Ib{\bm{I}}
\def\Pb{\bm{P}}

\def\thetab{\bm{\theta}}
\def\betab{\bm{\beta}}

\def\d#1{\,\mbox{d}#1}
\def\argmins#1#2{\mbox{arg}\min_{#1}\left\{{#2}\right\}}
\def\argmaxs#1#2{\mbox{arg}\max_{#1}\left\{{#2}\right\}}
\def\argmin#1#2{\mathop{\mbox{arg}\min}_{#1}\left\{{#2}\right\}}
\def\argmax#1#2{\mathop{\mbox{arg}\max}_{#1}\left\{{#2}\right\}}
\def\expf#1{\mbox{exp}\left\{{#1}\right\}}
\def\intg{\int\kern-1.1em\int}

\def\trace{\mbox{trace}}

\def\xmap{\widehat{\xb}_{\mbox{\tiny MAP}}}
\def\xmmap{\widehat{\xb}_{\mbox{\tiny MMAP}}}
\def\zerob{\bm{0}}

\def\thetah{\widehat{\theta}}
\def\betah{\widehat{\beta}}

\def\xbh{\widehat{\xb}}
\def\thetabh{\widehat{\thetab}}
\def\betabh{\widehat{\betab}}

\def\xbhk{\widehat{\xb}^{k}}
\def\thetahk{\widehat{\theta}^{k}}
\def\betahk{\widehat{\beta}^{k}}

\def\xbhkp{\widehat{\xb}^{k+1}}
\def\thetahkp{\widehat{\theta}^{k+1}}
\def\betahkp{\widehat{\beta}^{k+1}}

\def\thetabhk{\widehat{\thetab}^{k}}
\def\betabhk{\widehat{\betab}^{k}}

\def\thetabhkp{\widehat{\thetab}^{k+1}}
\def\betabhkp{\widehat{\betab}^{k+1}}

\def\lra{\longrightarrow}
\def\esp#1{\mbox{E}\left\{ #1 \right\}}

\def\var#1{\mbox{Var}\left\{#1\right\}}
\def\intd{\int\kern-.8em\int}
\begin{document}
\maketitle
\thispagestyle{empty}

\begin{abstract}
The main object of this paper is to present some general concepts of Bayesian 
inference and more specifically the estimation of the hyperparameters in inverse 
problems. We consider a general linear situation where we are given
some data $\yb$ related to the unknown parameters $\xb$ by 
$\yb=\Ab \xb+\nb$ and  where we can assign the 
probability laws $p(\xb|\thetab)$, 
$p(\yb|\xb,\betab)$, $p(\betab)$  and $p(\thetab)$.  
The main discussion is then how to infer $\xb$, $\thetab$ and $\betab$ 
either individually or any combinations of them. 
Different situations are considered and discussed. 
As an important example, we consider the case where $\theta$ and $\beta$ 
are the precision parameters of the Gaussian laws to whom we assign Gamma priors 
and we propose some new and practical algorithms to estimate them simultaneously. 
Comparisons and links with other classical methods such as maximum likelihood 
are presented. 
\end{abstract}


\section{Introduction}
In a general Bayesian inference, we have the data $\yb$, 
a known relation between the unknown parameters $\xb$ and $\yb$ 
and finally the hyperparameters $\betab$ 
and $\thetab$. 
The Bayesian estimation technique is now well established 
\cite{Box72,Sorenson80,Besag89,Green90,Malec92,Gindi93,Bernardo94}  
and has been used since many years to resolve the inverse problems in signal and 
image reconstruction and restoration 
\cite{Geman84,Tarantola87,Skilling88,Titterington85b,Demoment89a,Idier93,Djafari93b,Diebolt94,Carfantan95}.

The first step before applying the Bayes' rule is to 
assign the prior probability laws $p(\xb|\thetab)$, 
$p(\yb|\xb,\betab)$, $p(\thetab)$ and $p(\betab)$.  
The next step is to determine the posterior laws and then to infer the 
unknowns. In this paper we are focusing more on the second step than on the 
first step. So we assume that all the direct probability laws are known.  

The main object of this paper is to show how can we infer simultaneously 
the unknown parameters $\xb$ and the hyperparameters $\betab$ and $\thetab$ 
from the data $\yb$.  

Before going more in details let us give one example. This will permit us to fix 
the situations. 
Consider the case where the unknown parameters $\xb$ represent the 
pixel values of an unobserved image and the data $\yb$ are the pixel values of 
an observed image which is assumed to be a degraded version of it.  
If we consider a linear degradation we have
\begin{equation}
\yb=\Ab \xb+\nb,
\end{equation}
where $\Ab$ is a $(m\times n)$ matrix representing the degradation process 
and $\nb$ represents the measurement uncertainty (noise) which is assumed to 
be additive, centered, white, Gaussian and independent of $\xb$. 
This hypothesis leads us to
\begin{equation}
p(\yb|\xb,\beta)
=\frac{1}{Z_1(\beta)} \, \expf{-\frac{1}{2}\beta (\yb-\Ab\xb)^t (\yb-\Ab\xb)}.
\end{equation}
In this case $\beta$ is a positive parameter which is related to the noise 
variance $\sigma_b^2$ by $\beta=1/\sigma_b^2$ and 
$Z_1(\beta)=(2\beta/\pi)^{m/2}$ is the normalizing factor.

Consider also, for this example, a Gaussian prior law for $\xb$~:
\begin{equation}
p(\xb|\theta)=\frac{1}{Z_2(\theta)} \, \expf{-\frac{1}{2}\theta \phi(\xb)}
\quad\mbox{with}\quad 
\phi(\xb)=\xb^t \Pb_0^{-1} \xb,
\end{equation}
where $\theta=1/\sigma_x^2$ is a positive parameter, $\Pb_0$ is the 
{\em a priori} covariance matrix of $\xb$ and 
$Z_2(\theta)=(2\theta/\pi)^{n/2} |\Pb_0|^{1/2}$.

A well known case is the situation where $\theta$, $\beta$ and $\Pb_0$ 
are known and we only want to estimate $\xb$. In fact, in this special case, 
the joint law $p(\yb,\xb|\thetab,\betab)$ and the posterior law 
$p(\xb|\yb,\thetab,\betab)$ are both Gaussian and we have 
\begin{equation}
p(\xb|\yb,\thetab,\betab)
\propto \expf{-\frac{1}{2}\beta (\yb-\Ab\xb)^t (\yb-\Ab\xb)
-\frac{1}{2}\theta \xb^t \Pb_0^{-1} \xb},
\end{equation}
and, if we note by  
\begin{equation}
\xbh=\argmins{\xb}{J(\xb)
=(\yb-\Ab\xb)^t (\yb-\Ab\xb)-\lambda \xb^t \Pb_0^{-1} \xb} 
\hbox{~~with~~} \lambda=\theta/\beta,  
\end{equation}
then, it is easy to show that
\begin{equation}
\xb|\yb \sim {\cal N}\left(\xbh,\widehat{\Pb}\right) \hbox{~~with~~} 
\left\{\begin{array}{lcl}
\xbh &=& \beta \widehat{\Pb} \Ab^t \yb \\ 
\widehat{\Pb}  &=& \beta^{-1}\left(\Ab^t\Ab+\lambda \Pb_0^{-1}\right)^{-1}.
\end{array}\right.
\end{equation}
One can make a comparison with the classical regularization techniques for 
inverse problems with smoothness hypothesis, where $\Pb_0^{-1}=\Db^t \Db$ 
with $\Db$ a matrix approximating a differentiation operator and $\lambda$ 
is called the regularization parameter \cite{Demoment89a}.   

What we address here is the generalization of the problem of 
the determination of the regularization parameter $\lambda$ which has been 
studied for a long time 
\cite{Cullum79,Titterington85a,Younes88,Lakshmanan89,Djafari91a,Thompson91,Gassiat92,Liang92,Djafari93b,Bouman94,Iusem94} 
and is still an open problem. 

What is proposed here is to consider the general case where $\thetab$ 
and $\betab$ are considered to be unknown and we are facing to make inference 
as well about $\xb$ as about them. 
What we propose is to consider the hyperparameters 
$\thetab$ and $\betab$  in the same manner than $\xb$, {\em i.e}; 
translate our prior knowledge about them by the probability laws 
$p(\thetab)$ and $p(\betab)$, then determine the posterior laws and finally 
infer about them from these posterior laws. 

%In the following we consider the general case of the linear inverse problem 
%and just try to write formally how the Bayesian inference works. 

\section{General Bayesian inference approach} 

Assume now that we know the expressions of all the prior laws. 
We can then calculate the joint probability law:  
\begin{equation}
p(\yb,\xb,\thetab,\betab)
=p(\yb|\xb,\betab) \, p(\xb|\thetab) \, p(\thetab) \, p(\betab).
\end{equation}
In an ideal case where we are given $\Ab$, $\yb$, $\betab$ and $\thetab$, 
to infer $\xb$ we can calculate the posterior law $p(\xb|\yb,\thetab,\betab)$ 
and if we choose as the solution to our problem the Maximum {\em a posteriori} 
(MAP) estimate, we have:
\begin{equation}
\xbh
=\argmaxs{\xb}{p(\xb|\yb,\thetab,\betab)} 
=\argmaxs{\xb}{p(\yb|\xb,\betab) \, p(\xb|\thetab)}.
\end{equation}
But, unfortunately, in practical situations we are not given $\betab$ 
and $\thetab$ and the main problem is how to infer them. We consider the 
following situations:

\begin{enumerate} 
\item The first is to estimate the three quantities simultaneously. 
We call this method Joint Maximum {\em a posteriori} (JMAP) and the 
estimates are defined as
\begin{equation}
(\xbh,\thetabh,\betabh)
=\argmaxs{(\xb,\thetab,\betab)}{p(\yb|\xb,\betab) \, 
p(\xb|\thetab) \, p(\betab) \, p(\thetab)}.  
\end{equation}
One practical way to do this joint optimization is to use the following 
algorithm 
\begin{equation} 
\left\{\begin{array}{lcl}
\xbhkp    
&=&\displaystyle{\argmaxs{\xb}{p(\yb|\xb,\betabhk)\, p(\xb|\thetabhk)}} \\ 
\thetabhkp
&=&\displaystyle{\argmaxs{\thetab}{p(\xbhk|\thetab)\, p(\thetab)}} \\ 
\betabhkp
&=&\displaystyle{\argmaxs{\betab}{p(\yb|\xbhk,\betab)\, p(\betab)}}  
\end{array}\right.
\end{equation}

\item In the second case $\thetab$ and $\betab$ are considered 
as the nuisance parameters and are integrated out of the problem 
and $\xb$ is estimated by 
\begin{eqnarray}
\xbh
&=&\argmaxs{\xb}{p(\xb|\yb)}
=\argmaxs{\xb}{\intg p(\yb,\xb,\thetab,\betab)\d{\thetab}\d{\betab}} 
\nonumber\\ 
&=&\argmaxs{\xb}{\intg p(\yb|\xb,\betab) \, p(\betab) \d{\betab} 
                  \intg p(\xb|\thetab) \, p(\thetab) \d{\thetab}}. 
\end{eqnarray}
We call this method Marginalized MAP type one (MMAP$^1$).

\item In the third case only $\thetab$ is considered as the 
nuisance parameter and is integrated out of the problem and $\xb$ and 
$\betab$ are estimated by 
\begin{eqnarray}
(\xbh,\betabh)
&=&\argmaxs{(\xb,\betab)}{p(\xb,\betab|\yb,\thetab)} 
=\argmaxs{(\xb,\betab)}{\intg p(\yb,\xb,\thetab,\betab)\d{\thetab}} 
\nonumber\\ 
&=&\argmaxs{(\xb,\betab)}{p(\yb|\xb,\betab)\, p(\betab) 
\intg p(\xb|\thetab) \, p(\thetab) \d{\thetab}}.
\end{eqnarray}
We call this method Marginalized MAP type two (MMAP$^2$).

\item Finally, in the last case we may first estimate $\thetabh$ and 
$\betabh$ by 
\begin{eqnarray}
(\thetabh,\betabh)
&=&\argmaxs{(\thetab,\betab)}{p(\thetab,\betab|\yb)}  
=\argmaxs{(\thetab,\betab)}{\intg p(\yb,\xb,\thetab,\betab)\d{\xb}} 
\nonumber\\ 
&=&\argmaxs{(\thetab,\betab)}{
p(\betab) \, p(\thetab) \intg p(\yb|\xb,\betab)\, p(\xb|\thetab) \d{\xb}}
\nonumber\\ 
&=&\argmaxs{(\thetab,\betab)}{p(\betab)\, p(\thetab) l(\thetab,\betab|\yb)}.
\end{eqnarray}
and then used them for the estimation of $\xb$ by
\begin{eqnarray}
\xbh=\argmaxs{\xb}{p(\xb|\yb,\thetabh,\betabh)}.
\end{eqnarray}
We call this method Marginalized MAP type three (MMAP$^3$).

Note that if $p(\thetab)$ and $p(\betab)$ are uniform functions of 
$\thetab$ and $\betab$, then $\thetabh$ and $\betabh$ correspond to 
the classical maximum likelihood (ML) estimates because $l(\thetab,\betab|\yb)$ 
is, for a given $\yb$, the likelihood function of $\thetab$ and $\betab$. 

The calculus of $l(\thetab,\betab|\yb)$ is not easy and so is its optimization. 
Many works have been done on the subject. We distinguish three kind 
of methods: 

%\begin{itemize} 
-- The first is to use the Expectation-Maximization (EM) algorithm which has  
been developed exactly in the context of ML parameter estimation 
\cite{Dempster77,Green90,Vardi93}.  

-- The second is to estimate the integral using a Monte Carlo simulation method 
(Stochastic EM: SEM). 

-- The third is to make some approximations. 
For example, at each iteration during the optimization, one may obtain 
an analytical expression for that integral by approximating the 
expression inside it by a second order polynomial 
(Gaussian quadrature approximation). 
%\end{itemize} 
\end{enumerate}
We will consider this last method. 

\section{A case study}
Let us consider the following simple linear inverse problem 
$\yb=\Ab \xb+\nb$ 
and make the following hypothesis:

\begin{itemize}
\item The noise $\nb$ is considered to be white, centered and 
Gaussian with precision $\beta$, so that we have
\begin{equation}
\yb|\xb,\beta \sim {\cal N}(\Ab\xb,\beta^{-1}\Ib) \lra
p(\yb|\xb,\beta)
=\frac{1}{Z_1(\beta)} \, \expf{-\frac{1}{2} \beta\|\yb-\Ab\xb\|^2}.
\end{equation}
where $Z_1(\beta)\propto \beta^{m/2}$.

\item Our prior prior knowledge about $\xb$ can be translated by 
\begin{equation}
p(\xb|\theta)=\frac{1}{Z_2(\theta)} \, \expf{-\frac{1}{2} \theta\phi(\xb)}.
\end{equation}
where we will consider the following special cases for $\phi(\xb)$:
\begin{itemize}
\item Gaussian priors:
\[
\phi_G(\xb)=\xb^t\Pb_0^{-1}\xb=\|\Db\xb\|^2 \lra 
\xb|\theta \sim {\cal N}(\zerob,\theta^{-1}\Pb_0^{-1}),
\]
which can also be written 
\(\phi_G(\xb)=\sum_j\sum_i \, p_{ij}\, x_i x_j\)  
with some special cases: 
\[
\phi_{G}(\xb)=\sum_j \, x_j^2, \hbox{~~or~~} 
\phi_{G}(\xb)=\sum_j \, |x_j-x_{j-1}|^2.
\] 
\item Generalized Gaussian priors:
\[
\phi_{GG}(\xb)=\sum_j \, |x_j-x_{j-1}|^p,\quad  1<p \leq 2.
\]

\item Entropic priors:
\[
\phi_E(\xb)=\sum_{j=1}^n S(x_j) 
\hbox{~where~} 
S(x_j)=\left\{x_j^2,\,  x_j\ln x_j-x_j, \, \ln x_j-x_j\right\}.
\]
\item Markovian priors:
\[
\phi_M(\xb)=\sum_{j}\sum_{i\in N_j} V(x_j,x_i),   
\quad \hbox{where}\quad V(x_j,x_i) \hbox{~is a potential function }
\]
and where $N_j$ is a set of sites considered to be neighbors of site $j$, 
for example $N_j=\{j-1, j+1\}, \quad\hbox{or}\quad N_j=\{j-2, j-1, j+1, j+2\}$. 
\end{itemize}
Note that, in all cases $\theta$ is generally a positive parameter. 
Note also that in the first case we have $Z_2(\theta)\propto \theta^{n/2}$. 
Unfortunately we have not an analytic expression for $Z_2(\theta)$ in the 
other cases. 
However, in the situations we are concerned with, $Z_2(\theta)$ can either be 
calculated numerically or approximated by 
\begin{equation}
Z_2(\theta)\propto \theta^{\alpha n/2}.
\end{equation}

\item $\theta$ and $\beta$ are both positive parameters. 
We choose Gamma prior laws for them: 
\[
\theta \sim {\cal G}(a,\zeta)\lra
p(\theta)\propto \theta^{(a-1)}\expf{-\zeta\theta}
\lra \esp{\theta}=a/\zeta, \quad \var{\theta}=a/\zeta^2
\]
\[
\beta \sim {\cal G}(b,\zeta)\lra
p(\beta)\propto \beta^{(b-1)}\expf{-\zeta\beta}
\lra \esp{\beta}={b}/{\zeta}, \quad \var{\beta}={b}/{\zeta^2}
\]
\end{itemize}
Now, using the following notations
\[
Q(\xb)= \|\yb-\Ab\xb\|^2, \qquad 
J_0(\xb)=\beta Q(\xb)+\theta \phi(\xb),
\]
\[
\nabla Q(\xb)=- 2\Ab^t(\yb-\Ab\xb), \quad \hbox{and}\quad 
\nabla J_0(\xb)=\beta \nabla Q(\xb)+\theta \nabla \phi(\xb),
\]
we can calculate the expression of the joint $pdf$ 
$p(\yb,\xb,\theta,\beta)=p(\yb|\xb,\beta)\, p(\xb|\theta)\, p(\theta)\, p(\beta)$, 
which can be written 
\begin{equation}
p(\yb,\xb,\theta,\beta)
\propto
\theta^{-(\alpha n/2-a+1)} \beta^{-(m/2-b+1)}
\expf{- \frac{1}{2} J_1(\xb)},
\end{equation}
\begin{equation}
\hbox{with}\qquad J_1(\xb)=\beta[Q(\xb)+2\zeta]+\theta[\phi(\xb)+2\zeta]
=J_0(\xb)+2\zeta(\theta+\beta).
\end{equation}
This will let us to go further in details of some of the above mentioned cases. 
For example in the Gaussian case we have:
\[
\xb|\yb,\theta,\beta 
\sim  {\cal N}(\xbh,\widehat{\Pb})
\hbox{~with~} \xbh=\beta\left(\beta\Ab^t\Ab+\theta \Pb_0^{-1}\right)^{-1}\Ab^t\yb
\hbox{~and~} \widehat{\Pb}=\left(\beta\Ab^t\Ab+\theta\Pb_0^{-1}\right)^{-1}
\]
\[
\begin{array}{l}
\displaystyle{
\theta|\yb,\xb,\beta 
\sim  {\cal G}(a-\alpha n/2,\frac{1}{2}[\phi(\xb)+2\zeta])
\lra\esp{\theta|\yb,\xb,\beta}=\frac{2a-\alpha n}{[\phi(\xb)+2\zeta]}},
\\ 
\displaystyle{
\beta|\yb,\xb,\theta 
\sim  {\cal G}(b-m/2,\frac{1}{2}[Q(\xb)+2\zeta])
\lra\esp{\beta|\yb,\xb,\theta}=\frac{2b-m}{[Q(\xb)+2\zeta]}.
} 
\end{array}
\]
Now, let us consider the four aforementioned methods a little more in 
details. 

\subsection{Joint Maximum A Posteriori ($\mbox{JMAP}$)}
Using the expressions and the notations of the last paragraph 
in (11) we have to deal with the following algorithm:
\begin{eqnarray*}
\xbhkp    
&=&\argmaxs{\xb}{p(\yb|\xb,\betahk) \, p(\xb|\thetahk)}  
=\argmins{\xb}{J_0(\xb,\betahk,\thetahk)}, \\ 
\thetahkp
&=&\argmaxs{\theta}{p(\xbhk|\theta)\,p(\theta)}  
=\argmins{\theta}{[\phi(\xbhk)+2\zeta] \theta -(2a-\alpha n-2)\ln\theta}, \\ 
\betahkp
&=&\argmaxs{\beta}{p(\yb|\xbhk,\beta)\,p(\beta)}   
=\argmins{\beta}{[Q(\xbhk)+2\zeta] \beta -(2b-m-2)\ln\beta}.
\end{eqnarray*}
The two last equations have explicit solutions. 
In the case of Gaussian priors, the first equation has also an explicit solution. 
However, in general, we propose the following gradient based algorithm:
\begin{eqnarray*}
\hbox{\bf Algorithm 1:}\quad 
\xbhkp    
&=& (1-\mu)\xbhk-\mu\nabla J_0(\xbhk,\betahk,\thetahk) \\ 
&=&  (1-\mu)\xbhk-\mu[\betahk\nabla Q(\xbhk)+\thetahk\nabla \phi(\xbhk)]
,\quad  0<\mu<1, \\ 
\thetahkp
&=&\frac{(2a-\alpha n-2)}{[\phi(\xbhk)+2\zeta]}, \quad a>(\alpha n+2)/2, \\ 
\betahkp
&=&\frac{(2b-m-2)}{[Q(\xbhk)+2\zeta]}, \quad b>(m+2)/2.
\end{eqnarray*}
The conditions $a>(\alpha n+2)/2$ and $b>(m+2)/2$ are added to satisfy, 
when necessary, the positivity constraint of $\thetabh$ and $\betabh$.

\subsection{Marginalized Maximum A Posteriori $\mbox{MMAP}^1$}
Considering $\theta$ and $\beta$ as the nuisance parameters and integrating out 
them from $p(\yb,\xb,\theta,\beta)$ we obtain 
\begin{equation}
p(\yb,\xb)=\intd p(\yb,\xb,\theta,\beta)\d{\beta}\d{\theta}
\propto
[Q(\xb)+2\zeta]^{-(m-2b)/2} \, [\phi(\xb)+2\zeta]^{-(\alpha n-2a)/2}
\end{equation}
Now, defining \quad 
\(
\displaystyle{
\xmmap=\argmaxs{\xb}{p(\xb|\yb)}=\argmins{\xb}{\frac{1}{2} J_2(\xb)}
},
\)
\begin{equation}
\hbox{with}\qquad 
J_2(\xb)=(2a-\alpha n)\ln [Q(\xb)+2\zeta] +(2b-m) \ln [\phi(\xb)+2\zeta],
\end{equation}
and trying to calculate this solution by an iterative gradient based algorithm, 
we have to calculate 
\[
\nabla J_2(\xb)=\frac{(2a-\alpha n)}{[Q(\xb)+2\zeta]} \nabla Q(\xb)
                 +\frac{(2b-m)}{[\phi(\xb)+2\zeta]} \nabla \phi(\xb).
\]
We propose then the following iterative algorithm:
\begin{eqnarray*}
\hbox{\bf Algorithm 2:}\quad  
\xbhkp
&=& (1-\mu)\xbhk-\mu \nabla J_2(\xbhk) \\ 
&=& (1-\mu)\xbhk-\mu[\betahk\nabla Q(\xbhk)+\thetahk\nabla \phi(\xbhk)]
,\quad  0<\mu<1, \\  
\thetahk&=&\frac{(2a-\alpha n)}{[\phi(\xbhk)+2\zeta]}, \quad a>\alpha n/2, \\ 
\betahk&=&\frac{(2b-m)}{[Q(\xbhk)+2\zeta]}, \quad b>m/2.
\end{eqnarray*}
%which can be compared to the algorithm proposed in preceding section. 

\subsection{Marginalized Maximum A Posteriori $\mbox{MMAP}^2$}
In this case, $\theta$ only is considered as a nuisance parameter 
and is integrated out:
\begin{eqnarray}
p(\yb,\xb,\beta)
&=& \intg p(\yb,\xb,\theta,\beta)\d{\theta} \nonumber \\ 
&\propto&
\beta^{-m/2+b-1} [\phi(\xb)+2\zeta]^{-(\alpha n-2a)/2} 
\expf{-\frac{1}{2} \beta [Q(\xb)+2\zeta]}. 
\end{eqnarray}
Then, $\xb$ and $\beta$ are estimated by
\begin{equation}
(\xbh,\betah)=\argmax{\xb,\beta}{p(\yb,\xb,\beta)}.
\end{equation}
Noting 
\[
-2\ln p(\yb,\xb,\beta)=
-(2b-m-2)\ln\beta+(2a-\alpha n) \ln [\phi(\xb)+2\zeta]+\beta [Q(\xb)+2\zeta]
\]
and differentiating it with respect to $\beta$ gives
\[
\betah=\frac{2b-m-2}{[Q(\xb)+2\zeta]}. 
\]
So, noting 
\begin{equation}
J_3(\xb,\beta)=(2a-\alpha n) \ln [\phi(\xb)+2\zeta]+\beta [Q(\xb)+2\zeta]
\end{equation}
\[
\hbox{and}\qquad 
\nabla J_3(\xb,\beta)
=\frac{2a-\alpha n}{[\phi(\xb)+2\zeta]} \nabla\phi(\xb)+\beta\nabla Q(\xb), 
\]
and using a gradient based algorithm for minimizing $J_3$ with respect to $\xb$ 
we propose the following:
\begin{eqnarray*}
\hbox{\bf Algorithm 3:}\qquad   
\xbhkp   &=& (1-\mu)\xbhkp - \betahk \nabla Q(\xb) - \thetahk \nabla\phi(\xb) 
,\quad  0<\mu<1, \\  
\thetahk &=& \frac{2a-\alpha n}{[\phi(\xbhk)+2\zeta]}, \quad a>(\alpha n+2)/2 \\ 
\betahk  &=& \frac{2b-m-2}{[Q(\xbhk)+2\zeta]}, \quad b>(m+2)/2. 
\end{eqnarray*}

\subsection{Maximum Likelihood or $\mbox{MMAP}^3$}
In this case first $\xb$ integrated out $\xb$ from 
$p(\yb,\xb,\theta,\beta)$ to obtain:
\begin{equation}
p(\yb,\theta,\beta)
=\intg p(\yb,\xb,\theta,\beta)\d{\xb}
=\frac{\beta^{(b-1)}}{Z_2(\beta)}\frac{\theta^{(a-1)}}{Z_1(\theta)} 
\intg \expf{-\frac{1}{2} J_1(\xb,\beta,\theta)} \d{\xb}
\end{equation}
\begin{equation}
\hbox{with}\qquad\qquad
J_1(\xb,\beta,\theta)=\beta[Q(\xb)+2\zeta]+\theta[\phi(\xb)+2\zeta]. 
\end{equation}
Excepted the Gaussian case where $J_1$ is a quadratic function of $\xb$, 
in general, it is not easy to obtain an analytical 
expression for this integral. One can then try to make a Gaussian 
approximation which means to develop $J_1$ around its minimum 
\(
\xmap=\argmin{\xb}{J_1(\xb,\beta,\theta)}
\) 
by
\begin{equation}
J_1(\xb,\beta,\theta)
\simeq \frac{1}{2}(\xb-\xmap)^t \Mb (\xb-\xmap)+\gb^t (\xb-\xmap)+c, 
\end{equation}
where $\gb=\beta\nabla Q(\xb)+\theta\nabla \phi(\xb)$ is the gradient 
of $J_1$ and $\Mb$ is its Hessian, both calculated for $\xmap$. 
With this approximation we obtain 
\begin{equation}
p(\yb,\theta,\beta)
=\beta^{-m/2+b-1} \theta^{-\alpha n/2+a-1} 
|\Mb(\beta,\theta)|^{-\frac{1}{2}} \expf{-\frac{1}{2} J_1(\xmap,\beta,\theta)}. 
\end{equation}
Differentiating $l(\theta,\beta|\yb)=\ln p(\yb,\theta,\beta)$ 
with respect to $\beta$ and $\theta$ 
gives
\[
\betah=\frac{2b-m-2}{[Q(\xbhk)+2\zeta]+\trace[\Mb^{-1}\Ab^t\Ab]}, \quad 
\thetah=\frac{2a-\alpha n-2}{[\phi(\xbhk)+2\zeta]+\trace[\Mb^{-1}\Pb_0^{-1}]}.  
\]
where $\Pb_0^{-1}$ is the Hessian of $\phi(\xb)$, $\Ab^t\Ab$ is the Hessian of 
$Q(\xb)$ and $\Mb$ is the Hessian of $J_1(\xb)$:
\[
\Mb(\beta,\theta)=\beta\Ab^t\Ab+\theta\Pb_0^{-1}.
\]
Using these expressions we propose the following algorithm:
\begin{eqnarray*}
\hbox{\bf Algorithm 4:}\qquad  
\xbhk&=&\argmin{\xb}{J_1(\xb,\betahk,\thetahk)}
       =\Mb(\betahk,\thetahk)^{-1}\Ab^t\yb, \\ 
\thetahkp&=&\frac{2a-\alpha n-2}{[\phi(\xbhk)+2\zeta]+\trace[\Mb^{-1}\Pb_0^{-1}]}, \\  
\betahkp&=&\frac{2b-m-2}{[Q(\xbhk)+2\zeta]+\trace[\Mb^{-1}\Ab^t\Ab]}. 
\end{eqnarray*}
This algorithm needs the inversion of the matrix $\Mb$ which is very costly 
in practice. 

\section{Comparison and the main structure of the proposed algorithmes}
Comparing the {\bf Algorithms 1 to 4}, one can see that they all have the 
same structure: 
\begin{itemize}
\item for fixed $\theta$ and $\beta$ optimize locally a criterion 
$J(\xb,\beta,\theta)$, and    
\item update $\theta$ and $\beta$ using the solution $\xbh$ just obtained 
and iterate until convergence.
\end{itemize}
Note also that only in {\bf Algorithm 4}, the updating step takes account 
of the measurement system operator $\Ab$ and the covariance structure 
$\Pb_0$ of the input $\xb$. 

\section{Conclusions and perspectives}
We considered the inverse problem of infering the unknowns $\xb$ from 
the data $\yb$ in a special case of linear inverse problems 
$\yb=\Ab \xb+\nb$ using a full bayesian approach and presented four 
algorithms to estimate simultanously the hyperparameters $\theta$ 
and $\beta$ and the unknowns $\xb$. 
The main structure of all of these algorithms are the same even if the 
procedure to deduce them have been different. 
However, we have not yet really tested them to give any conclusion about 
their relative performances. 
Note however that one of them distinguishes itself from the others by 
taking account of the measurement system operator $\Ab$ and the covariance 
structure $\Pb_0$ of $\xb$ in the hyperparameters updating step and, 
by the same way, by its calculation cost. 
We hope to be able to give some measure of their relative performances 
in simulation and in real applications in near future. 

%My definitions
%--------------------------------
%\input{revuedef.tex}
%\bibliographystyle{maxent95}
%\bibliography{/users/brutus/djafari/Tex/Inputs/gpibase,/users/brutus/djafari/Tex/Inputs/amd,/users/brutus/djafari/Tex/Inputs/herve,/users/brutus/djafari/Tex/Inputs/mila}
%\bibliography{gpibase,amd,herve,mila}

\begin{thebibliography}{10}
\bibitem{Malec92}
D.~Malec and J.~Sedransk, ``Bayesian methodology for combining the results from
  different experiments when the specifications for pooling are uncertain,''
  {\em \BMK}, vol.~79, no.~3, pp.~593--601, 1992.
\end{thebibliography}

\def\AA{Astrononmy and Astrophysics}
\def\ABE{Annals of Biomedical Engineering}
\def\AMS{Annals of mathematical Statistics}
\def\AISM{Annals of Institute of Statistical Mathematics}
\def\AO{Applied Optics}
\def\AP{The Annals of Probability}
\def\AST{The Annals of Statistics}
\def\BMK{Biometrika}
\def\CPAM{Communications on Pure and Applied Mathematics}
\def\EMK{Econometrica}
\def\CRAS{Compte-rendus de l'acad\'emie des sciences}
\def\CVGIP{Computer Vision and Graphics and Image Processing}
\def\GJRAS{Geophysical Journal of the Royal Astrononomical Society}
\def\GSC{Geoscience}
\def\GPH{Geophysics}
\def\GRETSI{Actes du GRETSI} 
\def\CGIP{Computer Graphics and Image Processing}
\def\ICASSP{Proceedings of IEEE ICASSP}
\def\ieeP{Proceedings of the IEE}
\def\ieeeAC{IEEE Transactions on Automatic and Control}
\def\ieeeAES{IEEE Transactions on Aerospace and Electronic Systems}
\def\ieeeAP{IEEE Transactions on Antennas and Propagation}
\def\ieeeASSP{IEEE Transactions on Acoustics Speech and Signal Processing}
\def\ieeeBME{IEEE Transactions on Biomedical Engineering}
\def\ieeeCS{IEEE Transactions on Circuits and Systems}
\def\ieeeCT{IEEE Transactions on Circuit Theory}
\def\ieeeC{IEEE Transactions on Communications}
\def\ieeeGE{IEEE Transactions on Geoscience and Remote Sensing}
\def\ieeeGEE{IEEE Transactions on Geosciences Electronics}
\def\ieeeIP{IEEE Transactions on Image Processing}
\def\ieeeIT{IEEE Transactions on Information Theory}
\def\ieeeMI{IEEE Transactions on Medical Imaging}
\def\ieeeMTT{IEEE Transactions on Microwave Theory and Technology}
\def\ieeeM{IEEE Transactions on Magnetics}
\def\ieeeNS{IEEE Transactions on Nuclear Sciences}
\def\ieeePAMI{IEEE Transactions on Pattern Analysis and Machine Intelligence}
\def\ieeeP{Proceedings of the IEEE}
\def\ieeeRS{IEEE Transactions on Radio Science}
\def\ieeeSMC{IEEE Transactions on Systems, Man and Cybernetics}
\def\ieeeSP{IEEE Transactions on Signal Processing}
\def\ieeeSSC{IEEE Transactions on Systems Science and Cybernetics}
\def\ieeeSU{IEEE Transactions on Sonics and Ultrasonics}
\def\ieeeUFFC{IEEE Transactions on Ultrasonics Ferroelectrics and Frequency Control}
\def\IJC{International Journal of Control}
\def\IJCV{International Journal of Computer Vision}
\def\IJIST{International Journal of Imaging Systems and Technology}
\def\IP{Inverse Problems}
\def\IUSS{Proceedings of International Ultrasonics Symposium}
\def\JAPH{Journal of Applied Physics}
\def\JAP{Journal of Applied Probability}
\def\JAS{Journal of Applied Statistics}
\def\JASA{Journal of Acoustical Society America}
\def\JASAS{Journal of American Statistical Association}
\def\JCAM{Journal of Computational and Applied Mathematics}	% Ajout HC
\def\JCAT{Journal of Computer Assisted Tomography}
\def\JEWA{Journal of Electromagnetic Waves and Applications}	% Ajout HC
\def\JMO{Journal of Modern Optics}
\def\JNDE{Journal of Nondestructive Evaluation}		% Ajout HC
\def\JMP{Journal of Mathematical Physics}
\def\JOSA{Journal of Optical Society America}
\def\JRSSA{Journal of Royal Statistical Society A}
\def\JRSSB{Journal of Royal Statistical Society B}
\def\JRSSC{Journal of Royal Statistical Society C}
\def\JTSA{Journal of Time Series Analysis}              %Ajout MF
\def\JVCIR{Journal of Visual Communication and Image Representation} 
	\def\MMAS{???} % Trouve dans gpi base 
\def\MNAS{Mathematical Methods in Applied Science}
\def\MNRAS{Monthly Notes of Royal Astronomical Society}
\def\MP{Mathematical Programming}
\def\OC{Optics Communication}
\def\PRA{Physical Review A}
\def\PRB{Physical Review B}
\def\PRC{Physical Review C}
\def\PRL{Physical Review Letters}			% Ajout HC	
\def\RGSP{Review of Geophysics and Space Physics}	% Ajout HC			% Ajout HC	
\def\RS{Radio Science}					% Ajout HC	
\def\SP{Signal Processing}
\def\siamAM{SIAM Journal of Applied Mathematics}
\def\siamCO{SIAM Journal of Control and Optimization}
\def\siamJO{SIAM Journal of Optimization}		% Ajout HC pour JFB
\def\siamMA{SIAM Journal of Mathematical Analysis}
\def\siamNA{SIAM Journal of Numerical Analysis}
\def\siamO{SIAM Journal of Optimization}
\def\siamR{SIAM Review}
\def\TMK{Technometrics}
\def\TS{Traitement du Signal}
\def\UMB{Ultrasound in Medecine and Biology}
\def\US{Ultrasonics}
\def\USI{Ultrasonic Imaging}
%--------------------------------

\end{document}